\documentclass[12pt]{iopart}
\usepackage{graphicx}
\usepackage{color}

\begin{document}

\title{$\mathrm{CO_2}$ exploding clusters dynamics probed by XUV fluorescence}

\author{M. Negro$^1$, H. Ruf$^2$, B. Fabre$^2$, F. Dorchies$^2$, M. Devetta$^1$, D. Staedter$^3$, C. Vozzi$^1$, Y. Mairesse$^2$, S. Stagira$^{4}$}

\address{$^1$CNR-IFN, I-20133 Milan, Italy}
\address{$^2$CELIA, Universit\'{e} de Bordeaux, CEA, CNRS, F-33405 Talence, France}
\address{$^3$Universit\'{e} de Toulouse, UPS, F-31062 Toulouse, France}
\address{$^4$Politecnico di Milano - Physics Department, I-20133 Milan, Italy}

\ead{salvatore.stagira@polimi.it}

\begin{abstract}
Clusters excited by intense laser pulses are a unique source of warm dense matter, that has been the subject of intensive experimental studies. The majority of those investigations concerns atomic clusters, whereas the evolution of molecular clusters excited by intense laser pulses is less explored. In this work we trace the dynamics of $\mathrm{CO_2}$ clusters triggered by a few-cycle 1.45-$\mu$m driving pulse through the detection of XUV fluorescence induced by a delayed 800-nm ignition pulse. Striking differences among fluorescence dynamics from different ionic species are observed.
\end{abstract}

\maketitle

\section{Introduction}
The nonlinear interaction of laser pulses with clusters has been the subject of intense theoretical and experimental research in the last two decades \cite{Fennel_2010}. Particular attention has been devoted to the complex cluster dynamics occurring during and after the interaction with a high-energy ultrashort laser pulse; time-resolved experiments have been performed in single and double-pulse configurations, looking to electron and ion spectra \cite{Snyder_1996,Springate_2000,Fukuda_2003}, to the efficiency in laser absorption and scattering \cite{Zweiback_1999,Zweiback_2000} as well as to the emission in the XUV \cite{Parra_2000,Mori_2001,Chen_2002,Lin_2004} and X ray \cite{Parra_2003,Issac_2004,Dorchies_2005} spectral regions.\\
\begin{figure}[b]
 \begin{center}
 \includegraphics[width=0.8\textwidth]{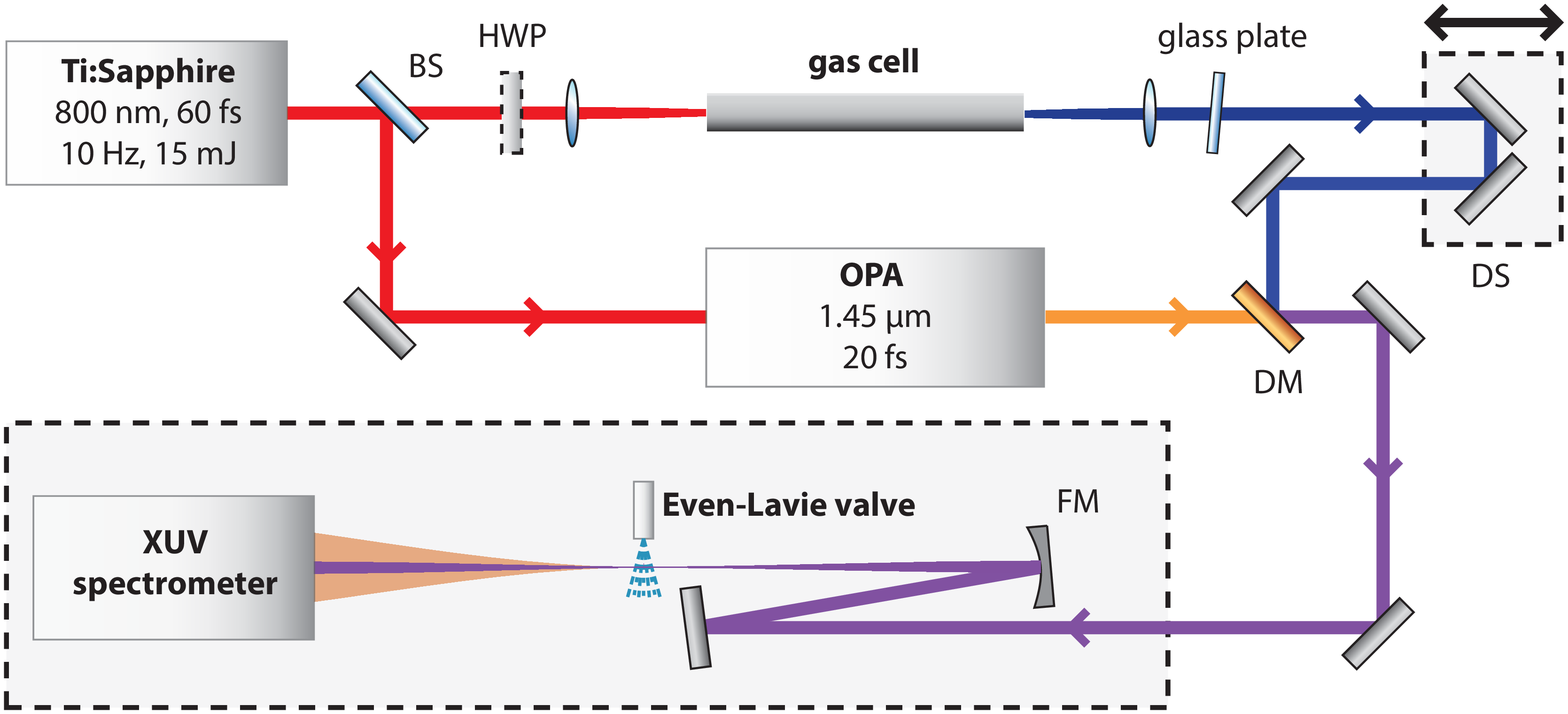}
 \end{center}
 \centering
\caption{Experimental setup. BS: beam splitter; HWP: half-wave plate; OPA: optical parametric amplifier; DS: delay stage; DM: dichroic mirror; FM: focusing mirror. \label{setup}}
\end{figure} 
Most of these time-resolved studies have been performed with Ti:Sapphire lasers operating at 800 nm, which were the only sources of intense and ultrashort laser pulses in operation up to a decade ago. However impressive advances in laser science are quickly driving the investigation of laser-clusters interaction towards new frontiers. On the one hand, studies concerning clusters exposed to intense XUV an X ray pulses are nowadays made possible by the availability of Free Electron Lasers \cite{Bostedt_2008,Iwayama_2012,Iwan_2012,Thomas_2012} and laser-driven High Order Harmonic sources \cite{Murphy_2008,Hoffmann_2011}. On the other hand, the recent development of intense and ultrafast laser sources based on optical parametric amplifiers \cite{Vozzi_2007} has opened the way to the study of strong-field laser-matter interaction in the mid-IR spectral region \cite{Vozzi_2012}. In spite of the large number of experimental investigations, the mentioned studies mostly concentrate on atomic clusters made of noble 
gases, whereas the evolution of molecular clusters excited by intense laser pulses is much less explored \cite{Snyder_1996,Iwan_2012}.\\
In this work we present an experimental investigation of $\mathrm{CO_2}$ cluster dynamics driven by a few-cycle 1.45-$\mu$m laser pulse and probed by a delayed 800-nm pulse; the dynamics was traced by recording XUV fluorescence lines emitted by different ionic species as a function of the pump-probe delay and of the average cluster size. Fluorescence is a powerful probe of the cluster environment, since it is sensitive to the electron temperature as well as to collisional excitation and ionization mechanisms at work. Its potential can be fully accomplished by time-resolved techniques, as in a pump-probe measurement, providing access to dynamical processes occurring in the cluster.
The main findings of our investigation are that: \textit{(i)} at a larger cluster size corresponds a slower growing and decay dynamics of XUV ionic emission lines; \textit{(ii)} the fluorescence decay dynamics is well fitted by a linear decay law; \textit{(iii)} lower ionization stages emit earlier than higher ones along the pump-probe delay scan. The latter finding points to intriguing opportunities for probing phenomena occurring inside clusters excited by intense laser pulses.\\ 
By a simple elaboration of the experimental outcomes we also evaluate the initial electron temperature of the $\mathrm{CO_2}$ clusters ionized by the driver pulse; we show that, in spite of the relatively high peak intensity of the driver, the interaction of a $\mathrm{CO_2}$ cluster with a mid-IR pulse leads to the generation of a cold nanoplasma, with a temperature inversely proportional to the square of the number of molecules inside the cluster. This result shows that intense mid-IR pulses, which are usually considered as ideal drivers for accelerating electrons to very high ponderomotive energies \cite{Vozzi_2012}, can be effectively exploited as ultrafast drivers of nanoscale warm matter with solid-state density and electron temperatures in the range of a few to a few tens of electronvolts.

\begin{table}[t]
  \centering
 \begin{tabular}{ccc}
 \hline \hline
 $p$ (bar) & $T$ ($^\circ$C) & $N$ (molecules)\\ [0.5ex]   
 \hline
  12 & 80 & 3200\\
  12 & 31.4 & 6100\\
  30 & 150 & 8200\\
  30 &  80 & 17000\\
  30 & 33.7 & 31000\\  [1ex] 
  \hline\hline
  \end{tabular} 
  \caption{Experimental conditions explored in this work; $p$: $\mathrm{CO_2}$ valve backing pressure; $T$: valve temperature; the corresponding average cluster sizes $N$ were calculated according to \cite{Hagena_1992}.}
  \label{tab:sizes}
 \end{table} 

\section{Experimental setup}
The experimental setup is shown in Figure \ref{setup}. A Ti:Sapphire amplified laser system provides 60-fs pulses at 10 Hz repetition rate centered at 800 nm with an energy of 15 mJ. A large portion of the beam (about 90\% in energy) pumps an Optical Parametric Amplifier (OPA) which provides tunable, intense and ultrashort pulses in the mid-IR \cite{Vozzi_2007}. The OPA can operate between 1.3 and 1.8 $\mu$m, generating pulses with a temporal duration between 18 and 25 fs, according to the working wavelength, and mJ-level energy. In the present work, the OPA was tuned at 1.45 $\mu$m and produced a 1.3 mJ pump pulse with 20 fs duration (hereafter called \textit{driver}).\\
The remaining portion of the Ti:Sapphire beam (about 10\% in energy) is used for generating the probe pulse (hereafter called \textit{heater}). This beam undergoes optical filamentation in an argon-filled cell (not shown in the figure); the filamentation process ensures a good spatial profile of the beam at the output of the cell. Furthermore, the large spectral broadening induced by filamentation is exploited for temporal stretching of the pulse up to 100 fs by propagation through a glass plate. Afterwards, the heater beam propagates through a half-wave plate for controlling the polarization state.\\
The driver and the heater are recombined with a dicroic beam splitter and propagate through an iris which is used for tuning the on-target energy. A translation stage is used for changing the temporal delay between the pulses; in the following, positive delays correspond to a heater pulse coming after the driver one. The beam is then sent in the vacuum interaction chamber where a spherical mirror (curvature radius $R=-300$ mm) focuses it into the target. The on-target pulse energies were 0.45 mJ (driver) and 0.3 mJ (heater); the estimated peak intensities were about $2 \times 10^{14} \ \mathrm{W/cm^2}$ and $4 \times 10^{13} \ \mathrm{W/cm^2}$ respectively. In this work the target was a jet of clusters obtained by supersonic expansion of noble or molecular gases \cite{Hagena_1992} through an Even-Lavie valve \cite{Luria_2011} with a trumpet shaped nozzle of 250 $\mu$m diameter synchronized to the laser source and operating with an opening time of 25 $\mu$s. The average size of the clusters could be changed by tuning the gas 
backing pressure and the valve temperature; the experimental condition we explored are summarized in Table \ref{tab:sizes}.\\
\begin{figure}[t]
 \begin{center}
 \includegraphics[width=0.8\textwidth]{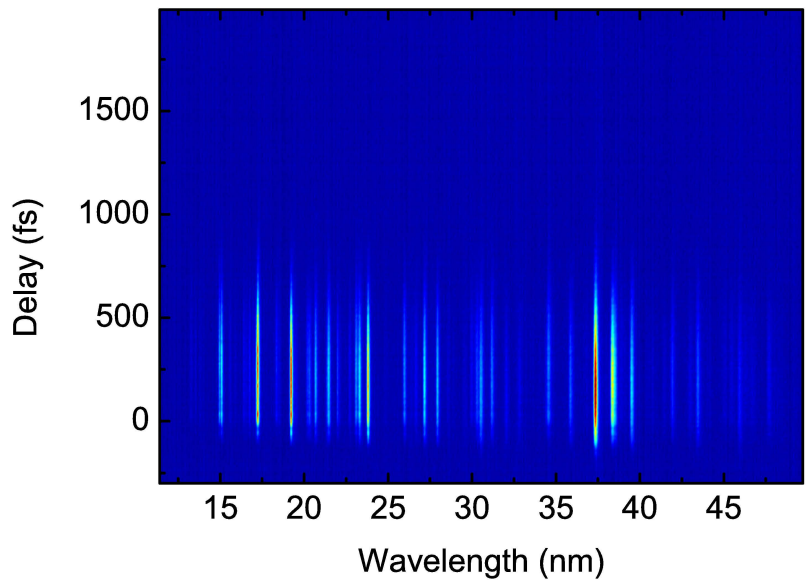}
 \end{center}
 \centering
\caption{Pump-probe scan in $\mathrm{CO_2}$ clusters as a function of the delay between driver and heater and of the fluorescence wavelength; the average number of molecules per cluster was $N=6100$. The driver and heater polarizations were perpendicular.\label{pump-probe}}
\end{figure} 
The XUV fluorescence emitted by the excited $\mathrm{CO_2}$ clusters was detected by a flat-field XUV spectrometer in grazing incidence geometry \cite{Poletto_2001} as a function of the delay between the two laser pulses; the spectrometer was calibrated in wavelength on the basis of some strong fluorescence lines emitted by Krypton clusters compared against the NIST Atomic Spectra Database \cite{NIST_2012}. The spectrometer was aligned on the direction of the laser beam, thus allowing the detection of coherent XUV radiation emitted by the excited clusters. For driver and heater with parallel polarizations, we observed continuous XUV spectra emitted in a very limited delay range, compatible with the duration of the pulses cross-correlation. This emission was interpreted as enhanced high-order harmonics generated when the two pulses are overlapped in time \cite{Vozzi_2009} and was exploited for determining the zero delay between the pulses. We could not observe generation of high-order harmonics in $\mathrm{CO_2}$ clusters neither 
by one of the two pulses nor when the pulses were not overlapped in time. Moreover, harmonic generation was never observed at any time delay for driver and heater pulses with orthogonal polarizations.\\
Since we will focus this work on incoherent emission from ions excited inside the clusters, we will consider hereafter only those experimental results we obtained with pulses having perpendicular polarization directions.

\section{Results}
A typical pump-probe scan performed in $\mathrm{CO_2}$ clusters is shown in Figure \ref{pump-probe} as a function of temporal delay and fluorescence wavelength. According to the Hagena model of clustering in supersonic gas expansion \cite{Hagena_1992} the average number of molecules per cluster was $N=6100$ (see Table \ref{tab:sizes} for the corresponding experimental conditions).\\ 
\begin{figure}[t]
 \begin{center}
 \includegraphics[width=0.8\textwidth]{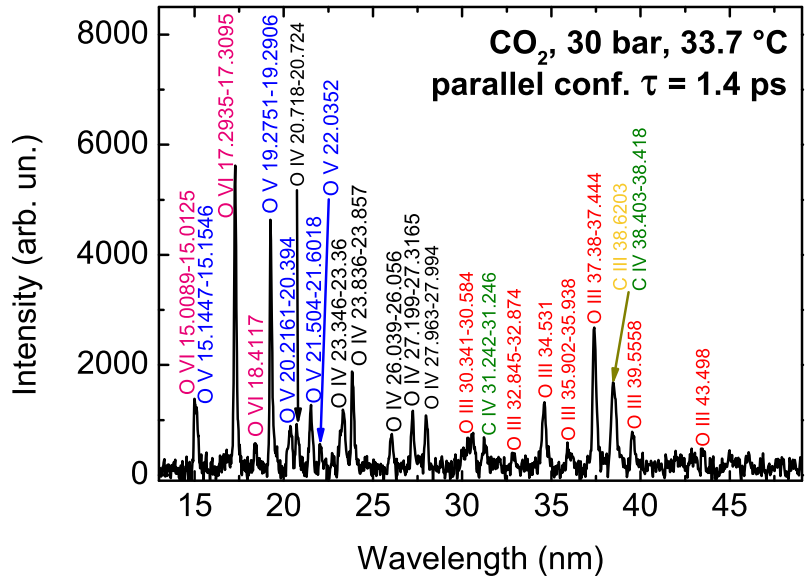}
 \end{center}
 \centering
\caption{Emission spectrum in $\mathrm{CO_2}$ clusters with average size $N=31000$ at a driver-heater delay $\tau=1.4$ ps; the driver and heater polarizations were parallel. Each fluorescence line is associated to the emitting ionic species and the peak wavelength. \label{identification}}
\end{figure}
Several emission lines are clearly observed; the lines appear only for positive delays, i.e. when the heater pulse comes after the driver one. This behaviour is consistent with the so called \textit{nanoplasma model} \cite{Ditmire_1996}, which provides a qualitative description of the laser-cluster interaction: the driving pulse prepares the cluster in a mild ionization state by optically-induced electron tunneling; afterwards, the ionized cluster starts expanding until it interacts with the heater pulse. Provided that the cluster plasma density is in a proper range (which corresponds to an optimal delay range), a very efficient absorption of the heater pulse takes place. During such interaction, efficient collisional ionization occurs and the new ionic species can readily be transferred to an excited state, from which they decay to the ground state by fluorescence emission.\\
It is worth to stress that the time-resolved experiment we performed must not be interpreted as a usual pump-probe measurement, in which the probe pulse does not perturb the excited state induced by the pump one in the sample. On the contrary, here the heater pulse ignites the clusters that are pre-ionized by the driver one, hence it strongly interacts with the sample.\\
Another important aspect to be considered is that the ionic fluorescence is induced by the heater, but evolves on temporal scales much longer than its duration since it involves ionization and excitation by thermal electrons as well as radiative electron-ion recombination processes. Hence the observed spectra are the integrated emission over all the temporal evolution of the excited plasma. This aspect must be kept in mind when considering the measurements we report in this work: when we consider a time-resolved fluorescence dynamics, we refer to the dependence of the fluorescence signal on the driver-heater delay, not to the duration of the fluorescence emission. 
 \begin{table}[t]
  \centering
 \begin{tabular}{lrcrr}
 \hline \hline
 Species & $I_p$ (eV) & Investigated lines (nm) &  Lower level & Upper level\\ [0.5ex]   
 \hline
  C I & 11.26 & & & \\
  O I & 13.61 & & & \\
  C II & 24.38 & & & \\
  O II & 35.12 & & & \\
  C III & 47.88 & 38.62(*) & $\mathrm{2s^2 \ ^1S_0}$ & $\mathrm{2s 3p \ ^1P_1}$\\
  O III & 54.93 & 37.38-37.444 & $\mathrm{2s^2 2p^2 \ ^3P_j}$ & $\mathrm{2s^2 2p 3s \ ^3P_{j'}}$\\
  C IV &  64.49 & 38.403-38.418(*) & $\mathrm{1s^2 2p \ ^2P_j}$ & $\mathrm{1s^2 3d \ ^2D_{j'}}$ \\
  O IV & 77.41 & 23.836-23.857 & $\mathrm{2s^2 2p \ ^2P_j}$ & $\mathrm{2s^2 3d \ ^2D_{j'}}$ \\
  O V & 113.90 & 19.2751-19.2906 & $\mathrm{2s 2p \ ^3P_j}$ & $\mathrm{2s 3d \ ^3D_{j'}}$ \\
  O VI & 138.12 & 17.2935-17.3095 & $\mathrm{1s^2 2p \ ^2P_j}$ & $\mathrm{1s^2 3d \ ^2D_{j'}}$\\
  C V & 392.09 & & & \\
  C VI & 489.99 & & & \\
  O VII & 739.33 & & &\\
  O VIII & 871.41 & & &\\  [1ex] 
  \hline\hline
  \end{tabular} 
  \caption{Ionization potential $I_p$, investigated spectral lines and corresponding transitions for Carbon and Oxygen ions \cite{NIST_2012}; data are ordered for increasing $I_p$. The lines from Carbon, marked with (*), appear overlapped in our setup.}
  \label{tab:ionpot}
 \end{table}
 
 \begin{figure}[t]
 \begin{center}
 \includegraphics[width=0.8\textwidth]{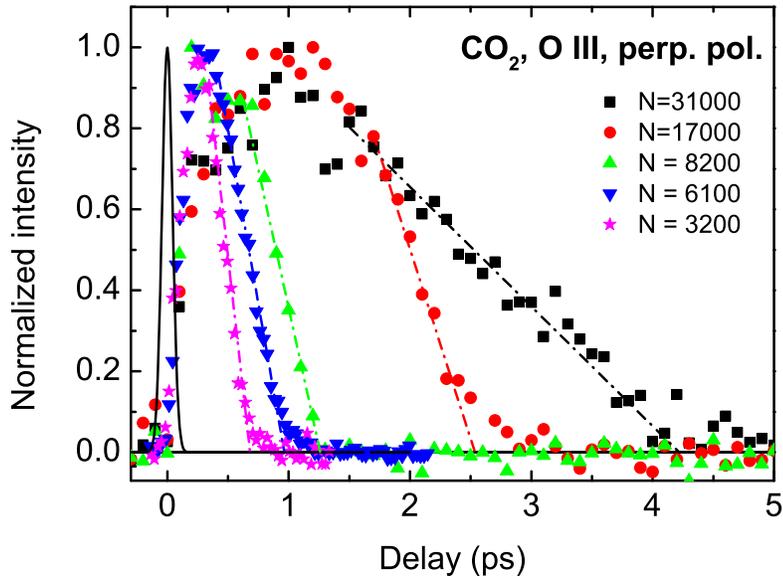}
 \end{center}
 \centering
\caption{Normalized fluorescence traces of the O III ion as a function of the delay between driver and heater, for different $\mathrm{CO_2}$ cluster sizes: $N=3200$ (stars), $N=6100$ (down triangles), $N=8200$ (up triangles), $N=17000$ (filled dots), $N=31000$ (filled squares); the dash-dotted lines are guides to the eye showing the almost linear decay of the ionic fluorescence. The driver and heater polarizations were perpendicular; the cross-correlation between the pulses is shown as solid line.\label{comparison_OIII}}
\end{figure} 
\subsection{Assignment of emission lines} 
By comparing the emission spectra with the NIST Atomic Spectra Database \cite{NIST_2012}, it is possible to assign each fluorescence line to the corresponding ionic emitting species and to the involved radiative transition. Figure \ref{identification} shows the identification of the most intense lines we observed; the experimental conditions are reported in the caption. Several emitting species are identified, in particular C III, C IV, O III, O IV, O V and O VI. The presence of relatively high ionization stages is not surprising by itself, since it is a fingerprint of laser-cluster interaction \cite{Fennel_2010}. However it is instructive to recall in Table \ref{tab:ionpot} the ionization potentials that must be overcome in order to observe such ions. On the basis of those parameters and of the Ammosov-Delone-Krainov model \cite{Ammosov_1986}, it is found for instance that isolated O VI ions could be generated by tunnel ionization of O V ions by exploiting driving pulses with a peak intensity three order of 
magnitude larger than the experimental one. This striking result reveals the strong importance of the cluster environment on the ionization processes during and after the interaction with the laser pulses.\\
Molecular clusters, as the ones considered in this work, are more complex systems than atomic clusters. For instance in carbon dioxide several molecular orbitals lie in the energy region between 13.78 eV and 22 eV under the vacuum level, hence electrons colliding with moderate kinetic energy on $\mathrm{CO_2}$ molecules could trigger several phenomena, like the generation of molecular ions and the molecular fragmentation along several pathways. Those phenomena may play important roles in the early stages of the cluster excitation. Since fluorescence from excited $\mathrm{CO_2}$ molecules and molecular fragments does not appear in the spectral region we considered, we will limit the investigation to the fate of atomic ions that are likely to be generated after complete molecular fragmentation. In the following we will refer to some of the observed fluorescence lines; the last three columns in Table \ref{tab:ionpot} report the spectral range, the lower and the upper levels of the transitions we will 
investigate 
hereafter. For each of the chosen set of lines, the temporal evolution of the fluorescence signal is obtained as the integral of the emission in the corresponding spectral range. It is worth noting that emission from C III and C IV ions cannot be disentangled owing to the limited spectral resolution of our setup. 

\begin{figure}[t]
 \begin{center}
 \includegraphics[width=0.7\textwidth]{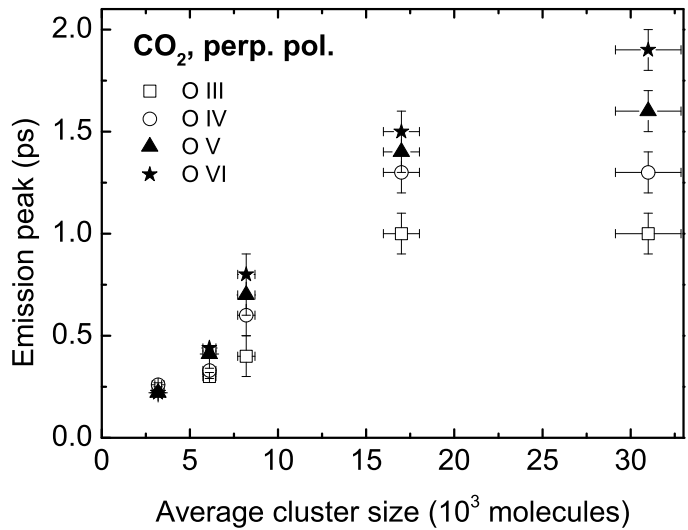}
 \end{center}
 \centering
\caption{Driver-heater delay corresponding to the peak of the ion emission as a function of the cluster size $N$ for the following ions: O III (empty squares), O IV (empty dots), O V (filled triangles) and O VI (filled stars). \label{peak_O}}
\end{figure} 
\subsection{Size dependence of the fluorescence signal}
Once the principal ionic lines are identified, one can determine the fluorescence growing and decay dynamics for each ionic species as a function of the driver-heater delay. Figure \ref{comparison_OIII} shows the normalized fluorescence traces corresponding to the emission of the O III ion for different average cluster sizes. It is worth noting that the driver and heater durations are substantially shorter than the time scale of the observed dynamics, as can be seen by comparing the fluorescence signals (symbols) with the calculated pump-probe cross-correlation (solid line). The fluorescence dynamics on the leading edge of the pump-probe traces shows an initial fast rise, which then slows down for larger cluster sizes; also the decay of the fluorescence signal is slower for larger clusters. This trend is a general feature in the measured spectra, as it is observed also for the other ionic species (not shown). It is worth noting that the fluorescence decay can be very well fitted by a linear law for all the average cluster sizes, as shown by the dash-dotted lines reported in 
Figure \ref{comparison_OIII}.\\
\begin{figure}[t]
 \begin{center}
 \includegraphics[width=0.7\textwidth]{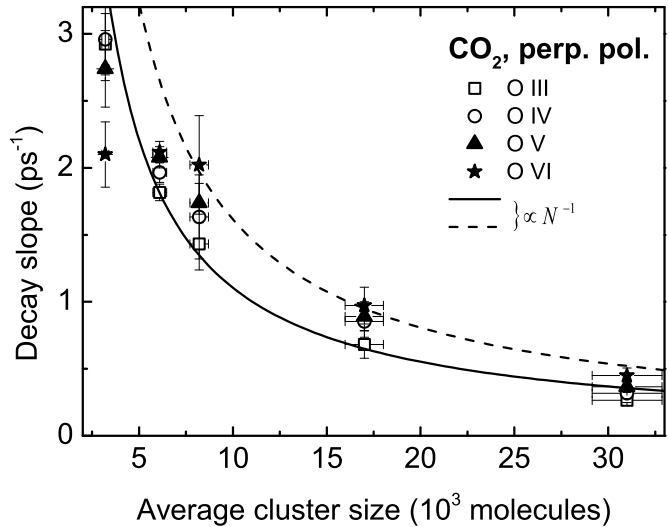}
 \end{center}
 \centering
\caption{Decay slope of the O III (empty squares), O IV (empty dots), O V (filled triangles) and of the O VI (filled stars) normalized fluorescence as a function of the cluster size $N$. The solid and dashed lines are fitting traces both proportional to $N^{-1}$. See text and Figure \ref{comparison_OIII} for more details. \label{slope_O}}
\end{figure} 
In order to provide a more quantitative description of the size dependence of the fluorescence dynamics, we report in Figure \ref{peak_O} the driver-heater delay at which the emission from each ionization stage is maximum; this optimal delay is shown as a function of the cluster size $N$. The error bars for the cluster sizes are chosen as equal to the relative width $\delta R/R $ of the size distribution in the cluster jet, which has been found in \cite{Dorchies_2003} to be of the order of 12\%; the error on the optimal emission delay has been determined as the sampling step used in the pump-probe measurement. As can be seen from the figure, in small clusters the emission peak occurs almost at the same delay for all the considered ionic species. This behavior is not observed in large clusters, where the fluorescence emitted by low ionization stages reaches the maximum much earlier than high stages; in particular in the largest clusters the optimal delay for O VI emission is almost double the 
one for O III 
emission.\\ 
As previously mentioned, the ionic fluorescence dynamics can be fitted with a linear temporal decay; Figure \ref{slope_O} shows the decay rates calculated for all the Oxygen ions; the error bars on the decay slope are determined from data fitting. The decay rates are inversely proportional to the cluster size, as confirmed by the $1/N$ curves plotted in the figure (solid and dashed lines, which respectively fit emission decay from O III and O VI ions, differing only by a constant multiplication factor). As a consequence, the cluster fluorescence persists for a delay range proportional to $N$, thus larger clusters emit for longer delay intervals. The agreement between measurement and fitting is fair for large cluster sizes, whereas becomes less satisfactory for small ones. In particular, the decay rate of the O VI fluorescence traces saturates to about 2.1 $\mathrm{ps^{-1}}$ in small clusters; such value is still well below the limit imposed by the pump-probe temporal resolution. A tendency in saturation, though less evident, is found for O V emission whereas is not observed for O III and O IV decay rates. On the basis of findings shown in Figures \ref{peak_O} and \ref{slope_O} we infer that the physical processes involved in ionic fluorescence emission depend on the cluster size, because one cannot reproduce all the emission processes with a unique scaling law. 
\begin{figure}[t]
 \begin{center}
 \includegraphics[width=0.8\textwidth]{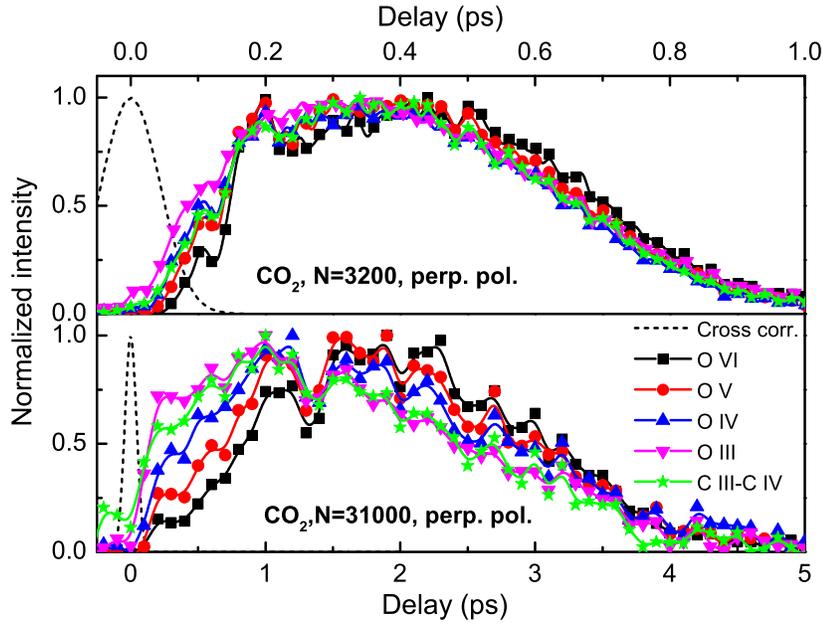}
 \end{center}
 \centering
\caption{Temporal evolution of the normalized fluorescence of O III (down triangles), O IV (up triangles), O V (filled dots) and O VI (filled squares) for an average cluster size $N=3200$ (upper panel) and $N=31000$ (lower panel); the fluorescence from C III and C IV is shown as stars. The driver and heater polarizations were perpendicular; the cross-correlation between the pulses is shown as dashed line. Solid lines are B-spline fittings of the data and are displayed only as a guide to the eye.\label{ion_31000_3200}}
\end{figure} 

\subsection{Dependence of the fluorescence signal on the ionic species}
As already mentioned in the previous section, the dynamics of ionic XUV fluorescence depends on the emitting species. In order to better clarify this peculiarity, we show in the lower panel of Figure \ref{ion_31000_3200} the temporal evolution of the normalized fluorescence for different ionic species and an average cluster size $N=31000$; symbols show the acquired data, whereas solid lines are B-spline fittings of the data and are reported only as a guide to the eye. The calculated cross-correlation between the two pulses is shown as dashed line.\\
Limiting initially the data analysis to Oxygen contributions, one finds that ionic species with lower ionization potential appear earlier than species with higher $I_p$ (see Table \ref{tab:ionpot} for comparison); moreover, the growing dynamics is steeper for the former species and smoother for the latter ones. As can be seen, the Carbon contribution seems to follow an intermediate dynamics between the O III and the O IV traces. Assuming that the ordering by ionization potential can be applied to Carbon as well, one can conclude that the line is dominated by the C IV contribution, at least for delays larger than 250 fs. It is worth noting that the ordering in fluorescence traces is confirmed for all the average cluster sizes we explored, although the overall growing dynamics becomes faster for smaller clusters. This trend can be seen for instance in the upper panel of Figure \ref{ion_31000_3200}, which shows the fluorescence dynamics for an average cluster size $N=3200$.\\
A close inspection to the fluorescence scan shows that a temporal ordering is also observed in the decay of the ionic emission, where emission from ions with higher $I_p$ seems to persist more than fluorescence emitted by ions with lower ionization potential. This ordering in the fluorescence decay was observed in the majority of cases but it was less evident for $N=17000$ and $N=6100$, probably owing to a larger noise amplitude. However, as will be clarified in the following, the departures in fluorescence decay among different ionic species is only apparent. This can be already understood from Figure \ref{slope_O}: for large clusters, the fluorescence decay rates behave according to the same scaling law as a function of cluster size (i.e. as $1/N$). Moreover all the traces show a linear temporal decay and, as can be seen from Figure  \ref{ion_31000_3200}, at a fixed cluster size the fluorescence emitted by all the ionic species goes to zero at the same delay. A consequence of this finding is that 
fluorescence decay traces from different ionic species can be rescaled in a suitable way in order to be overlapped and to show the same dynamics.

\section{Discussion} 
As already mentioned, in order to discuss the experimental data one has to keep in mind that the pump-probe measurement here presented are deeply in the non-perturbative regime. Hence a suitable way of analyzing the results is to divide the cluster dynamics in two stages: \textit{(a)} the first stage is triggered by the mid-IR driver pulse and consists in a mild ionization ad expansion of the clusters. The ionization and excitation rates can be assumed to be low since no ionic line emission was observed in the considered spectral range when the heater pulse was blocked. \textit{(b)} In the second stage, the heater pulse ignites the clusters, which undergo a complex dynamics giving rise to ionic fluorescence emission.\\
In the following, we will discuss these two stages separately.

\subsection{Cluster expansion}
\begin{figure}[t]
 \begin{center}
 \includegraphics[width=0.7\textwidth]{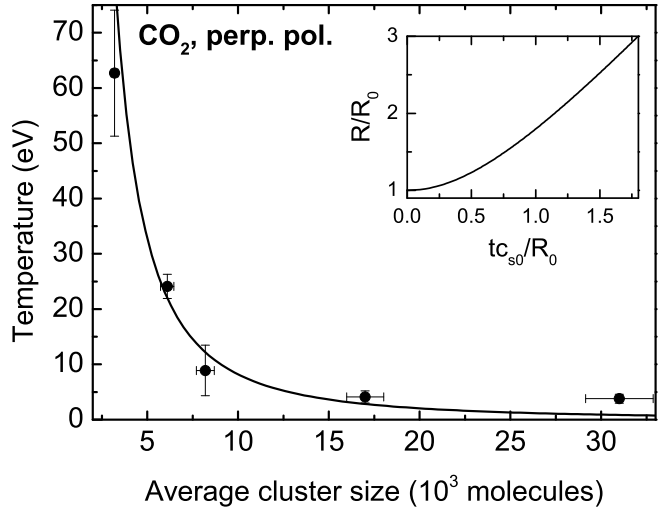}
 \end{center}
 \centering
\caption{Initial electronic temperature $ZT_{e0}$ of the clusters after driver excitation, calculated as a function of the cluster size $N$ (filled dots); the temperature follows a dependence on the cluster size as $N^{-2}$ (fit shown as solid line). Inset: evolution of the normalized cluster size $\tilde{R}=R/R_0$ as a function of the normalized time $\tilde{t}=t c_{s0}/R_0$ calculated in the framework of a self-similar expansion model \cite{Murakami_2006}. \label{Temp}}
\end{figure}

A long-standing debate in the interpretation of laser-cluster interaction concerns the mechanisms of cluster expansion; two main phenomena are in general considered: hydrodynamic expansion or Coulomb explosion \cite{Fennel_2010,Ditmire_1996}. The first mechanism is more likely to dominate in large clusters and for relatively cold electrons; in this case the freed electrons expand as a gas pulling ions outwards. Under suitable conditions the nanoplasma expansion can be represented by a self-similar model \cite{Murakami_2006,Beck_2009} where the key parameters governing the expansion are the initial electron temperature and the departure from charge neutrality inside the cluster. The second mechanism dominates in small clusters or for extremely high laser intensities and takes place when ions are not screened any more by the stripped electrons. In this case the repulsive forces dominate and the key parameter of the expansion process is the initial ion density and mean charge \cite{Grech_2011}. As a matter 
of fact, numerical simulations reveal that there is a faint border between the two mechanisms \cite{Jungreuthmayer_2004}. Moreover, when nonuniform expansion processes are considered, the cluster dynamics becomes extremely complicated since shock ion shells can be predicted \cite{Peano_2007} and have been also indirectly observed in the ion kynetic energy distribution of exploding clusters \cite{Erk_2011}. Further complications in the modelling of cluster dynamics come from transient phenomena like inner three-body recombination \cite{Thomas_2012} and macroscopic effects as the interaction among ions and Rydberg excited clusters \cite{Trivikram_2013}.\\
Even if the detailed theoretical study of $\mathrm{CO_2}$ cluster expansion is beyond the scope of this work, we note that in our experimental conditions a hydrodynamic expansion mechanism is likely to occur. In such a case, in the framework of a self-similar expansion model and assuming a uniform electron temperature inside the cluster, the characteristic size $R$ of the system should evolve in time according to the equation \cite{Murakami_2006}:
\begin{equation}
 \displaystyle \frac{d \tilde{R}}{d\tilde{t}} = 2  \sqrt{1-\tilde{R}^{-1}}
\end{equation} 
where $R_0$ is the initial size, $\tilde{R}=R/R_0$ is the normalized size and $\tilde{t}=t c_{s0}/R_0$ is the normalized time. The quantity $c_{s0}=\sqrt{ZkT_{e0}/m_i}$ is the ion sound speed in $t=0$, where $k$ is the Boltzmann constant, $Z$ the ionization stage, $T_{e0}$ the initial electron temperature and $m_i$ the ion mass. The solution of the equation takes the form:
\begin{equation}
 2 \tilde{t} = \sqrt{\tilde{R}\left(\tilde{R}-1 \right) } + \ln \left(\sqrt{\tilde{R}} + \sqrt{\tilde{R}-1} \right) 
\end{equation} 
The evolution of the normalized size is shown in the inset of Figure \ref{Temp}. Although self-similar solutions imply a non-uniform electron density inside the cluster, we will assume a constant density for the sake of discussion. One expects that the fluorescence peak is reached when the plasma density is about three times the critical density at the heater wavelength \cite{Ditmire_1996}; this allows one to estimate the order of magnitude of $ZT_{e0}$. In the case of $\mathrm{CO_2}$ clusters, we assume an average ion mass of 14.67 atomic units and an ion density at $t=0$ three times larger than the number of molecules per unit volume in the solid state. 
We took into account the peak of the O VI fluorescence traces as a reference (see Figure \ref{peak_O}) and we calculated the value of $ZT_{e0}$, which is reported in Figure \ref{Temp} as a function of the average cluster size $N$. One can see that for small clusters $ZT_{e0}$ is in the order of tens of eV but decreases suddenly to a few eV for large clusters; the estimated initial temperature depends on the cluster size according to $ZT_{e0} \propto N^{-2}$, as shown by the solid line in Figure \ref{Temp}. A small departure of the estimated $ZT_{e0}$ from the fitting is observed for $N=31000$, which could be attributed to a non-vanishing asymptotic value of the electron temperature for very large clusters.\\
The decrease in $ZT_{e0}$ for increasing cluster size is consistent with the findings of the \textit{nanoplasma model} \cite{Ditmire_1996}, which attributes this behavior to the more rapid expansion rate of smaller clusters that brings them near resonance earlier along the driver laser pulse. It is however important to stress that the the ionization stage $Z$ could be higher than 1 in small clusters owing to the mentioned resonance mechanism, hence for those clusters the effective electron temperature $T_{e0}$ could be lower than what is shown in Figure \ref{Temp}. Indeed this would also be consistent with the lack of ionic emission when the heater pulse was blocked, since too high values of the electron temperature would produce efficient ion impact excitation followed by radiative decay.\\
A noticeable consequence of the temperature estimation we report is that the assumption of a relatively cold electron gas in the nanoplasma is confirmed. It is worth stressing that such result is not trivial at first sight, since mid-IR driver pulses are expected to transfer a large ponderomotive energy to freed electrons with respect to standard 800-nm laser sources \cite{Vozzi_2012}. In particular, in our experimental conditions the maximum energy acquired by electrons ionized from a single molecule is expected to be in the order of 120 eV or more. However one has to recall that the physics of laser-cluster interaction strongly differs from the case of single and isolated molecules. The main heating mechanism occurring in laser-cluster interaction comes from field enhancement in the collisional nanoplasma. Since the cluster expands from overcritical densities towards lower ones, the resonance condition during the laser pulse is reached easier for short laser wavelenghts, which corresponds to higher 
critical 
densities. On the contrary, few-cycle mid-IR driving pulses leave the cluster in a colder state since this mechanism is quite inefficient.\\

\subsection{Cluster ignition}
The interaction of the expanding cluster with the heater pulse leads to strong energy absorption, which results in high collisional excitation and ionization rates and subsequent fluorescence emission. The way this energy is transferred to the nanoplasma is however still not completely understood. The model proposed by Ditmire et al. in 1996 assumed a uniform electron density and temperature inside the cluster during its expansion \cite{Ditmire_1996}, so that the resonance condition for efficient energy absorption could be reached only for a very limited amount of time, when the electron density reaches the condition $n_e \approx 3 n_c$ with $n_c$ the critical density at the laser wavelength. Milchberg et al. questioned this assumption and proposed that the non uniform plasma density in the cluster may play a key role in laser-cluster interaction \cite{Milchberg_2001}; moreover the cluster dynamics was found to be intrinsically bidimensional in nature even for spherical clusters, with further complications 
in the understanding of the energy coupling mechanisms \cite{Jungreuthmayer_2004,Milchberg_2001}. Recent theoretical models concerning ultracold expanding plasmas revealed that the absorption mechanisms depend on the departure of the cluster from neutrality \cite{Lyubonko_2012}. In the framework of a self-similar description, collective energy absorption is in general peaked inside the cluster where the laser frequency matches the local plasma frequency; however in charged clusters there is also a second absorption peak which is localized at the plasma edge and occurs at light frequencies related to the degree of charge imbalance \cite{Lyubonko_2012}.\\ 
\begin{figure}[t]
 \begin{center}
 \includegraphics[width=0.8\textwidth]{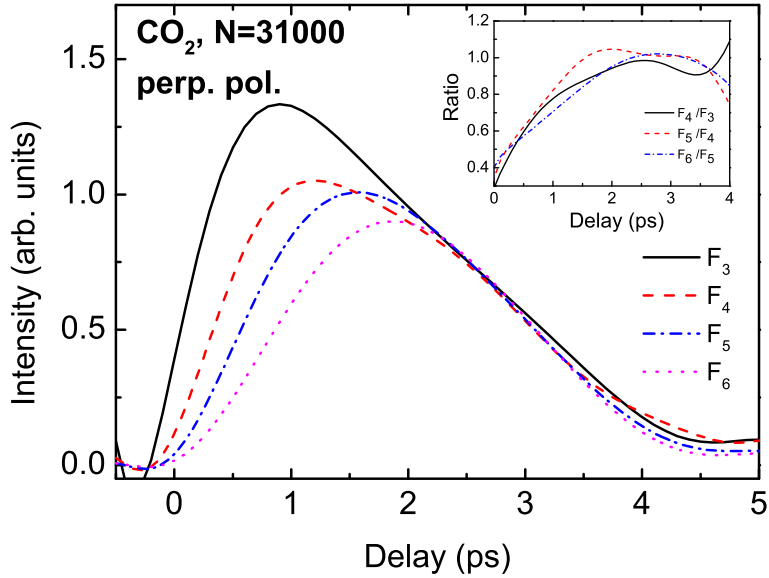}
 \end{center}
 \centering
\caption{Rescaled polynomial fittings of fluorescence traces already shown in the lower panel of Figure \ref{ion_31000_3200} (corresponding to $N=31000$ molecule per cluster) emitted by O III ($\mathrm{F_3}$, solid line), O IV ($\mathrm{F_4}$, dashed line), O V ($\mathrm{F_5}$, dash-dotted line) and O VI ($\mathrm{F_6}$, dotted line) ions. Inset: ratios between rescaled fittings; solid line: $\mathrm{F_{4}/F_{3}}$; dashed line: $\mathrm{F_{5}/F_{4}}$; dash-dotted line: $\mathrm{F_{6}/F_{5}}$. \label{fittings_O_31000}}
\end{figure} 
Although we are not aiming at a detailed understanding of cluster absorption mechanisms, we would like to exploit the experimental results for a qualitative description of the excitation and emission processes undergoing in the clusters. As already mentioned, in our experiments on $\mathrm{CO_2}$ clusters different ionization stages of Oxygen show different fluorescence temporal evolutions. In order to discuss in more detail such finding, we show in Figure \ref{fittings_O_31000} the polynomial fittings $F_i$ ($i=3,4,5,6$, corresponding respectively to fluorescence from O III, O VI, O V and O VI) of the traces shown in the lower panel of Figure \ref{ion_31000_3200} for $N=31000$ molecule per cluster. The fitting curves were rescaled in order to show that, in the delay range between 2 and 4 ps, the fluorescence presents the same dynamics irrespective of the emitting species. This is also shown in the inset of Figure \ref{fittings_O_31000}, where we plot the ratios among fitting curves corresponding to 
adjacent ionization stages; one can see that such ratios saturate at unity in the same delay range.\\
This finding allows us to speculate that, in a delay range comprised between 2 and 4 ps, the ratios among number of photons emitted from different ionic populations stay constant. 
On the other hand, for small driver-heater delays those ratios change and lower ionization stages always show higher $F_i$. Such behavior was observed for all the experimental conditions we analysed and it is consistent with the general dynamical scaling reported in Figure \ref{slope_O}. A tentative explanations for the observed dynamics can be the following: at small delays, when the cluster density is still overcritical, the heater absorption is moderate; for this reason the generation of high ionization stages is inefficient and low-charged ions predominate in fluorescence emission. Moreover electron-ion recombination would be at this stage very efficient, hence further reducing the contribution from highly charged ions.\\
At large delays almost all Oxygen atoms undergo efficient ionization up to O VI; these ions emit fluorescence but then undergo recombination with surrounding electrons. Such recombination is however less efficient with respect to small delays, owing to the lower electron and ion densities; hence the fluorescence we observe is a remnant of the various intermediate ionization stages, which have enough time for decaying to their groud state by photon emission before recombining with electrons. Thus at large pump-probe delays the fluorescence dynamics is mainly dictated by the dependence of O VI population on the delay.

\section*{Conclusions}
The interaction of clusters with intense and ultrashort laser pulses presents a very rich phenomenology. In this work we have shown that the fluorescence emitted by ions can be used as a sensitive probe of the expansion and excitation of molecular clusters. The efficient triggering of cluster expansion by 1.45-$\mu$m pulses demonstrate that ultrashort mid-IR pulses can be effectively used in this investigation. We found that fluorescence dynamics is a function of both the ionization stage and the cluster size. The scaling law we determined in the fluorescence traces points to a physical picture that cannot be easily related to simple models and calls for much more complex theories.\\
Our outcomes also demonstrate that mid-IR drivers leave the photoionized clusters with an almost solid-state density and a relatively cold electron temperature, allowing to reproduce warm and dense states of matter in a laser laboratory. The method we adopted for determining the initial temperature of the ionized cluster could be extended to study the effect of the driving wavelength on the cluster ionization and expansion rate and to fully determine the role of the ponderomotive energy and of the peak electric field on the initial stages of the laser-cluster interaction.

\ack
The research leading to the results presented in this work has received funding from LASERLAB-EUROPE (grant agreement n. 284464, EC's Seventh Framework Programme), from the European Research Council (ERC grant agreement n. 307964-UDYNI, EC's Seventh Framework Programme) and from the Italian Ministry of Research and Education (ELI project - ESFRI Roadmap).

\section*{References}

\bibliographystyle{iopart-num.bst}
\bibliography{clusters.bib}

\providecommand{\newblock}{}
\begin{thebibliography}{10}
\expandafter\ifx\csname url\endcsname\relax
  \def\url#1{{\tt #1}}\fi
\expandafter\ifx\csname urlprefix\endcsname\relax\def\urlprefix{URL }\fi
\providecommand{\eprint}[2][]{\url{#2}}

\bibitem{Fennel_2010}
Fennel T, Meiwes-Broer K~H, Tiggesbaumker J, Reinhard P~G, Dinh P~M and Suraud
  E 2010 {\em Reviews of Modern Physics\/} {\bf 82} 1793--1842

\bibitem{Snyder_1996}
Snyder E~M, Buzza S~A and Castleman A~W 1996 {\em Physical Review Letters\/}
  {\bf 77} 3347--3350

\bibitem{Springate_2000}
Springate E, Hay N, Tisch J~W~G, Mason M~B, Ditmire T, Marangos J~P and
  Hutchinson M~H~R 2000 {\em Physical Review A\/} {\bf 61} 044101

\bibitem{Fukuda_2003}
Fukuda Y, Yamakawa K, Akahane Y, Aoyama M, Inoue N, Ueda H and Kishimoto Y 2003
  {\em Physical Review A\/} {\bf 67} 061201(R)

\bibitem{Zweiback_1999}
Zweiback J, Ditmire T and Perry M~D 1999 {\em Physical Review A\/} {\bf 59}
  R3166--R3169

\bibitem{Zweiback_2000}
Zweiback J, Ditmire T and Perry M~D 2000 {\em Optics Express\/} {\bf 6}
  236--242

\bibitem{Parra_2000}
Parra E, Alexeev I, Fan J, Kim K~Y, McNaught S~J and Milchberg H~M 2000 {\em
  Physical Review E\/} {\bf 62} R5931--R5934

\bibitem{Mori_2001}
Mori M, Shiraishi T, Takahashi E, Suzuki H, Sharma L~B, Miura E and Kondo K
  2001 {\em Journal of Applied Physics\/} {\bf 90} 3595--3601

\bibitem{Chen_2002}
Chen L~M, Park J~J, Hong K~H, Choi I~W, Kim J~L, Zhang J and Nam C~H 2002 {\em
  Physics of Plasmas\/} {\bf 9} 3595--3599

\bibitem{Lin_2004}
Lin J~Y, Chu H~H, Shen M~Y, Xiao Y~F, Lee C~H, Chen S~Y and Wang J~P 2004 {\em
  Optics Communications\/} {\bf 231} 375--381

\bibitem{Parra_2003}
Parra E, Alexeev I, Fan J~Y, Kim K~Y, McNaught S~J and Milchberg H~M 2003 {\em
  Journal of the Optical Society of America B-Optical Physics\/} {\bf 20}
  118--124

\bibitem{Issac_2004}
Issac R~C, Vieux G, Ersfeld B, Brunetti E, Jamison S~P, Gallacher J, Clark D
  and Jaroszynski D~A 2004 {\em Physics of Plasmas\/} {\bf 11} 3491--3496

\bibitem{Dorchies_2005}
Dorchies F, Caillaud T, Blasco F, Bonte C, Jouin H, Micheau S, Pons B and
  Stevefelt J 2005 {\em Physical Review E\/} {\bf 71} 066410

\bibitem{Bostedt_2008}
Bostedt C, Thomas H, Hoener M, Eremina E, Fennel T, Meiwes-Broer K~H, Wabnitz
  H, Kuhlmann M, Pl{\"o}njes E, Tiedtke K, Treusch R, Feldhaus J, de~Castro
  A~R~B and M{\"o}ller T 2008 {\em Physical Review Letters\/} {\bf 100} 133401

\bibitem{Iwayama_2012}
Iwayama H, Nagasono M, Harries J~R and Shigemasa E 2012 {\em Optics Express\/}
  {\bf 20} 23174--23179

\bibitem{Iwan_2012}
Iwan B, Andreasson J, Bergh M, Schorb S, Thomas H, Rupp D, Gorkhover T, Adolph
  M, M{\"o}ller T, Bostedt C, Hajdu J and T\^{\i}mneanu N 2012 {\em Physical
  Review A\/} {\bf 86} 033201

\bibitem{Thomas_2012}
Thomas H, Helal A, Hoffmann K, Kandadai N, Keto J, Andreasson J, Iwan B,
  Seibert M, T\^{\i}mneanu N, Hajdu J, Adolph M, Gorkhover T, Rupp D, Schorb S,
  M{\"o}ller T, Doumy G, DiMauro L~F, Hoener M, Murphy B, Berrah N,
  Messerschmidt M, Bozek J, Bostedt C and Ditmire T 2012 {\em Physical Review
  Letters\/} {\bf 108} 133401

\bibitem{Murphy_2008}
Murphy B~F, Hoffmann K, Belolipetski A, Keto J and Ditmire T 2008 {\em Physical
  Review Letters\/} {\bf 101} 203401

\bibitem{Hoffmann_2011}
Hoffmann K, Murphy B, Kandadai N, Erk B, Helal A, Keto J and Ditmire T 2011
  {\em Physical Review A\/} {\bf 83}(4) 043203

\bibitem{Vozzi_2007}
Vozzi C, Calegari F, Benedetti E, Gasilov S, Sansone G, Cerullo G, Nisoli M,
  Silvestri S~D and Stagira S 2007 {\em Optics Letters\/} {\bf 32} 2957--2959

\bibitem{Vozzi_2012}
Vozzi C, Negro M and Stagira S {2012} {\em Journal of Modern Optics\/} {\bf
  {59}} {1283--1302}

\bibitem{Hagena_1992}
Hagena O~F 1992 {\em Review of Scientific Instruments\/} {\bf 63} 2374--2379

\bibitem{Luria_2011}
Luria K, Christen W and Even U 2011 {\em The Journal of Physical Chemistry A\/}
  {\bf 115} 7362--7367

\bibitem{Poletto_2001}
Poletto L, Tondello G and Villoresi P {2001} {\em {Review of Scientific
  Instruments}\/} {\bf {72}} {2868--2874}

\bibitem{NIST_2012}
Kramida A, Ralchenko Y, Reader J and Team N~A 2012 {\em NIST Atomic Spectra
  Database (ver. 5.0), [Online]\/} \urlprefix\url{http://physics.nist.gov/asd}

\bibitem{Vozzi_2009}
Vozzi C, Calegari F, Frassetto F, Poletto L, Sansone G, Villoresi P, Nisoli M,
  {De Silvestri} S and Stagira S {2009} {\em {Physical Review A}\/} {\bf {79}}
  033842

\bibitem{Ditmire_1996}
Ditmire T, Donnelly T, Rubenchik A~M, Falcone R~W and Perry M~D 1996 {\em
  Physical Review A\/} {\bf 53} 3379--3402

\bibitem{Ammosov_1986}
Ammosov M, Delone N and Krainov V {1986} {\em {Zhurnal Eksperimentalnoi I
  Teoreticheskoi Fiziki}\/} {\bf {91}} {2008--2013}

\bibitem{Dorchies_2003}
Dorchies F, Blasco F, Caillaud T, Stevefelt J, Stenz C, Boldarev A~S and
  Gasilov V~A 2003 {\em Physical Review A\/} {\bf 68} 023201

\bibitem{Murakami_2006}
Murakami M and Basko M~M 2006 {\em Physics of Plasmas (1994-present)\/} {\bf
  13} 012105

\bibitem{Beck_2009}
Beck A and Pantellini F 2009 {\em Plasma Physics and Controlled Fusion\/} {\bf
  51} 015004

\bibitem{Grech_2011}
Grech M, Nuter R, Mikaberidze A, {Di Cintio} P, Gremillet L, Lefebvre E,
  Saalmann U, Rost J~M and Skupin S 2011 {\em Physical Review E\/} {\bf 84}
  056404

\bibitem{Jungreuthmayer_2004}
Jungreuthmayer C, Geissler M, Zanghellini J and Brabec T 2004 {\em Physical
  Review Letters\/} {\bf 92} 133401

\bibitem{Peano_2007}
Peano F, Martins J~L, Fonseca R~A, Silva L~O, Coppa G, Peinetti F and Mulas R
  2007 {\em Physics of Plasmas (1994-present)\/} {\bf 14} 056704

\bibitem{Erk_2011}
Erk B, Hoffmann K, Kandadai N, Helal A, Keto J and Ditmire T 2011 {\em Physical
  Review A\/} {\bf 83} 043201

\bibitem{Trivikram_2013}
Trivikram T~M, Rajeev R, Rishad K~P~M, Jha J and Krishnamurthy M 2013 {\em
  Physical Review Letters\/} {\bf 111} 143401

\bibitem{Milchberg_2001}
Milchberg H~M, McNaught S~J and Parra E 2001 {\em Physical Review E\/} {\bf 64}
  056402

\bibitem{Lyubonko_2012}
Lyubonko A, Pohl T and Rost J~M 2012 {\em New Journal of Physics\/} {\bf 14}
  053039

\end{thebibliography}

\end{document}